\documentclass[a4paper,conference]{IEEEtran}
\usepackage{cmap}
\usepackage{indentfirst}
\usepackage{booktabs}
\usepackage{enumitem,amsmath,amssymb}
\usepackage{bm}
\usepackage{array}
\usepackage{multicol}
\usepackage{multirow}
\usepackage{graphicx}
\usepackage{color}
\usepackage{url}
\usepackage{cite}

\usepackage{placeins}
\usepackage[linesnumbered,ruled]{algorithm2e}
\ifCLASSOPTIONcompsoc
 \usepackage[caption=false,font=normalsize,labelfont=sf,textfont=footnotesize]{subfig}
\else
 \usepackage[caption=false,font=footnotesize]{subfig}
\fi

\begin{document}
\title{An Overlooked Role of Context-Sensitive Dendrites}

\author{Mohsin Raza$^{1,2}$, Ahsan Adeel$^{1,3*}$ \footnotemark}

\maketitle

\footnotetext{$^1$CMI Lab, University of Stirling, $^2$University of Wolverhampton, Wolverhampton. $^3$Oxford Computational Neuroscience, Nuffield Department of Surgical Sciences, University of Oxford, Oxford. \\ Email: ahsan.adeel@deepci.org}
\addtocounter{footnote}{-1}

\begin{abstract}
To date, most dendritic studies have predominantly focused on the apical zone of pyramidal two-point neurons (TPNs) receiving only feedback (FB) connections from higher perceptual layers and using them for learning. Recent cellular neurophysiology and computational neuroscience studies suggests that the apical input (context), coming from feedback and lateral connections, is multifaceted and far more diverse, with greater implications for ongoing learning and processing in the brain than previously realized. In addition to the FB, the apical tuft receives signals from neighboring cells of the same network as proximal (P) context, other parts of the brain as distal (D) context, and overall coherent information across the network as universal (U) context. The integrated context (C) amplifies and suppresses the transmission of coherent and conflicting feedforward (FF) signals, respectively. Specifically, we show that complex context-sensitive (CS)-TPNs flexibly integrate C moment-by-moment with the FF somatic current at the soma such that the somatic current is amplified when both feedforward (FF) and C are coherent; otherwise, it is attenuated. This generates the event only when the FF and C currents are coherent, which is then translated into a singlet or a burst based on the FB information. Spiking simulation results show that this flexible integration of somatic and contextual currents enables the propagation of more coherent signals (bursts), making learning faster with fewer neurons. Similar behavior is observed when this functioning is used in conventional artificial networks, where orders of magnitude fewer neurons are required to process vast amounts of heterogeneous real-world audio-visual (AV) data trained using backpropagation (BP). The computational findings presented here demonstrate the universality of CS-TPNs, suggesting a dendritic narrative that was previously overlooked.
\end{abstract}


\section{Introduction}
\label{s.intro}
Going beyond long-standing integrate-and-fire pyramidal Point Neurons (PNs) \cite{hausser2001synaptic}—upon which current deep learning is based \cite{lecun2015deep}—recent breakthroughs in cellular neurophysiology \cite{larkum1999new} have revealed that pyramidal neurons in the mammalian neocortex possess two major points of integration: apical and basal, termed TPNs. \cite{phillips2023cooperative}. To date, most studies, including the latest burst-dependent synaptic plasticity (BDSP) \cite{payeur2021burst} and single-phase deep learning in cortico-cortical networks \cite{Greedysingle}  have predominantly used FB information at the apical zone of TPNs for learning (aka. credit assignment)  \cite{lillicrap2020backpropagation, guerguiev2017towards,  sacramento2018dendritic, sarwat2022chalcogenide,zhang2021experimental,ward2022beyond}. However, the apical input, coming from the feedback and lateral connections, is far more diverse with far greater implications for ongoing learning and processing in the brain \cite{adeel2020conscious,adeel2022context, adeel2022unlocking}. The apical zone receives input from diverse cortical and subcortical sources as a context that selectively amplifies and suppresses the transmission of coherent and conflicting FF signals, respectively received at the basal zone \cite{adeel2020conscious,adeel2022context,phillips2017cognitive}. To uncover the true computational potential of TPNs, it is critical to emphasize the importance of understanding and defining the roles of different kinds of contexts arriving at the apical tuft \cite{adeel2020conscious}. Hence, dissection of contextual field (CF) into sub-CFs is imperative to better understand the amplification and suppression of relevant and irrelevant signals, respectively \cite{adeel2022context}. Specifically: (i) what kinds of information arrive at the apical tuft? (ii) how are they formed? (iii) how do they influence the cell’s response to the FF signals? \cite{adeel2022context}\\
The Conscious Multisensory Integration (CMI) theory \cite{adeel2020conscious, adeel2022context, adeel2022unlocking} suggested that the apical tuft of the TPN receives modulatory sensory signals coming from: the neighbouring cells of the same network e.g., audio as P, other parts of the brain, in principle from anywhere in space-time e.g., visuals as D, and the background information/overall coherent information across the multisensory (MS) network as U. These contextual signals play a decisive role in precisely selecting whether to amplify/suppress the transmission of relevant/irrelevant FF signals, without changing the content e.g., which information is worth paying more attention to? This, as opposed to, unconditional excitatory and inhibitory activity, is called conditional amplification and suppression \cite{adeel2020conscious}.\\
This view of context-sensitivity  is also called cooperative context-sensitive computing \cite{phillips2023cooperative, adeel2022context, adeel2022unlocking} whose processing and learning capabilities are shown to be well-matched to the capabilities of the mammalian neocortex \cite{sarwat2022chalcogenide, pagkalos2023leveraging}. In this approach, CS-TPNs receive P, D, and U fields to conditionally segregate relevant and irrelevant FF signals or transmit their FF message only when it is coherently related to the overall activity of the network. Individual neurons extract synergistic FF components as U by first conditionally segregating the coherent and incoherent multisensory information streams  and then recombining only the coherent multistream. The U is broadcasted to other brain areas which are received by other neurons along with the current local context (P and D) \cite{adeel2022context, muckli_2023_8380094, adeel2023unlocking}. These complex CS-TPNs when used in artificial neural network approaches enabled faster learning (beating Transformer—the backbone of ChatGPT) \cite{adeel2023cooperation} and consumed significantly fewer resources compared to convolutional neural nets (CNNs) \cite{adeel2022unlocking, adeel2022context} in some experimental settings. \\
Building on these recent findings, the goal of this work is to find the biologically plausible validation for why deep neural network (DNN) approaches with CS-TPNs processing and standard backpropagation-based learning \cite{adeel2022unlocking, adeel2022context, adeel2023cooperation}, manage to enable faster learning and consume fewer resources in some experimental settings compared to PNs-based DNNs. The main contributions of this paper are as follows:
\begin{enumerate}[label=\roman*.]
    \item We integrate the features of CMI-inspired CS TPNs \cite{adeel2020conscious, adeel2022context, adeel2022unlocking} into a spiking TPNs-inspired local BDSP rule \cite{payeur2021burst}, demonstrating accelerated `local' and `online' learning compared to BDSP approach alone. This validates the  efficient and effective information processing capabilities of  CS-TPNs, paving the way toward `local', `on-the-fly', `online' training and processing on neuromorphic chips.
\item Addressed the limitations raised in the BDSP rule \cite{payeur2021burst} by incorporating a thalamic circuitry proposed in \cite{adeel2020conscious, adeel2022context, aru2020cellular, adeel2024cellular} into our spiking model. The thalamic circuitry (termed as U) stores coherent information across the sensory hierarchy and extracts synergistic information. The thalamic inputs are projected to the apical dendrites of the CS-TPNs. We show one of the ways how U (as a Gate) can mediate the signalling between apical and basal inputs, depending on their strength, thereby selectively strengthening either short-term facilitation (STF) or short-term depression (STD). We show that the incorporation of this phenomenon, when integrated into a hierarchical circuit, helps the network learn faster.
\item We show that the spiking neural network composed of CS-TPNs significantly reduces the required number of events, including both singlets and bursts, for the task at hand compared to simple TPNs. 
\item We scaled up the two-layer CS TPNs-driven CNN \cite{adeel2022context, adeel2022unlocking} to a 50-layer deep CNN for AV speech processing, solely to demonstrate the scalability and information processing efficiency of CS-TPNs in larger networks. In some cases, it even surpasses the generalization capabilities of PNs-inspired CNNs. This shows the universality of our context-sensitive TPNs-inspired processing regardless of the learning mechanism. 
\end{enumerate}

\section{CMI-Inspired Cooperative CS TPNs + BDSP}
\begin{figure*} 
	\centering
    \includegraphics[trim=0cm 0cm 0cm 0cm, clip=true, width=\textwidth]{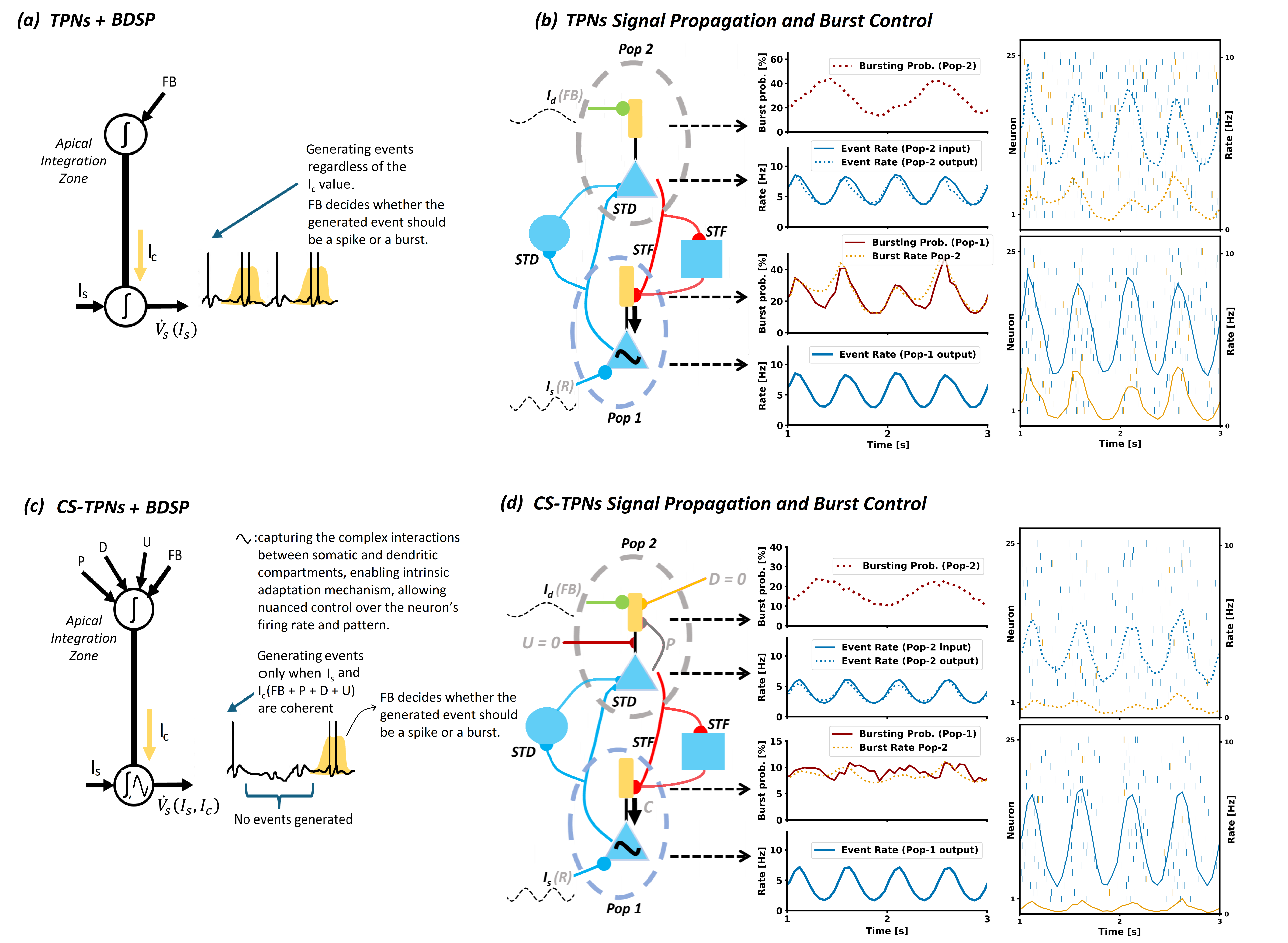}
	\caption{ Signal propagation and burst control in TPNs combined with short-term plasticity. Here, we show that the spiking neural network composed of CS-TPNs significantly reduces the required number of events, including both singlets and bursts, for the task at hand compared to simple TPNs. 
 \textbf{a} Context-insensitive TPNs with BDSP. The contextual current ($I_c$) does not influence the somatic potential; therefore, the learning is inspired by TPNs, but the processing is not, and is driven by point neuron conception. \textbf{(b)} Context-insensitive TPNs signal propagation and burst control \cite{payeur2021burst}. The circuit has two populations of neurons (Pop 1, bottom and Pop 2, top), 4000 neurons each. The neurons in Pop 1 receive external input as somatic current $I_s$ and those in Pop 2 receive dendritic current $I_d$. The feedforward pass (soma to soma) is from Pop 1 to Pop 2 through short-term depression (STD) synapses similar to \cite{payeur2021burst}. The output from Pop 1 is also projected to a population providing disynaptic inhibition (disk). The feed-backward pass (soma to dendritic compartment) is from Pop 2 to Pop 1 through the short-term facilitaion (STF) synapse. Similarly, the output from Pop 2 soma is connected to a population providing disynaptic inhibition (square). \textbf{c} Context-insensitive TPNs with BDSP, receiving FB as well as P, D, and U. Both learning and processing are influenced by the contextual current. The contextual current controls the membrane potential in a way that events occur only when the contextual current and somatic current are coherent. This is due to dense coordination and cooperation among TPNs, amplifying the transmission of coherent information and suppressing the transmission of incoherent information.  \textbf{(d)} Context-sensitive TPNs signal propagation and burst control. Neurons in both populations receive P, however, the D and U signals are set to zero for simplicity. The goal here is to test the context-sensitive operation and for that P is sufficient. Specifically, P is the contextual information coming from the neighbouring neurons within the same population. Please note that both populations have the same CS-TPNs dynamics. However, P, D, and U are included in the XOR simulations in Fig 2. It can be seen that the response of BP and ER at apical and soma, respectively, for both standard TPNs and CS-TPNs, after injecting $I_s$ in Pop 1, and $I_d$ in Pop 2 show different behaviour. Specifically, the CS-TPNs require fewer events, including singlets and bursts, to represent the input sinusoidal signal.}
\vspace{-1.2em}
\end{figure*}
\label{s.methodology}
\begin{figure*} 
	\centering
	\includegraphics[trim=0cm 0cm 0cm 0cm, clip=true, width=1\textwidth]{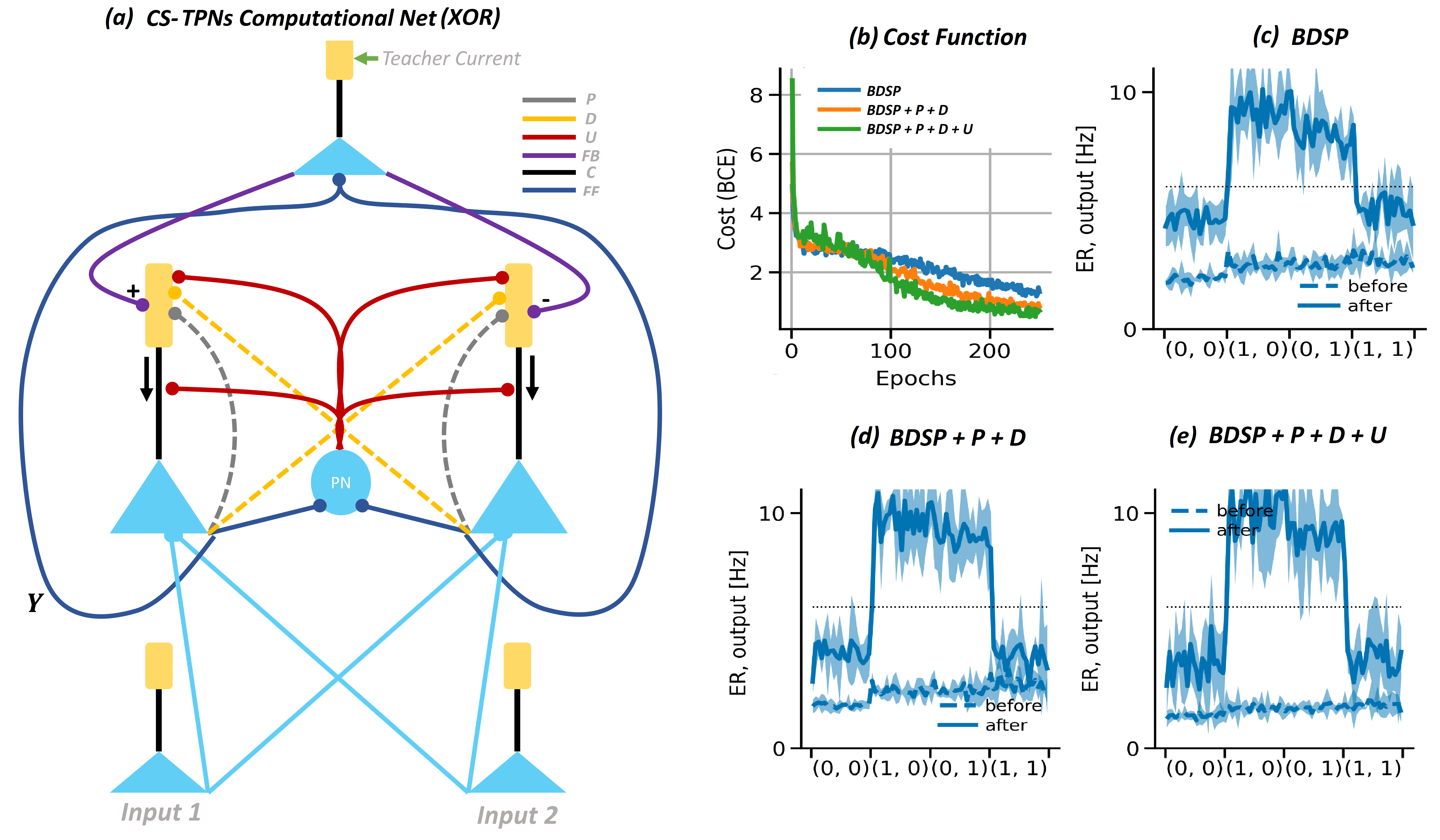}
	\caption{\textbf{(a)} CMI-inspired cooperative CS-TPNs + BDSP for XOR problem. Individual TPNs receive FB, P, D, and U inputs to conditionally segregate the coherent and incoherent FF signals, respectively. In terms of their sequence, first, coherent and incoherent signals are segregated by the TPNs. Then, these coherent signals are recombined by PNs, extracting synergistic FF components from all the coherent multistreams. This happens with the help of an additional ensemble representing U with a population of 50 PNs. U is broadcasted to TPNs along with the current local context \cite{adeel2020conscious, muckli_2023_8380094, adeel2023unlocking}. \textbf{(b)} Impact of this information processing mechanism: an increased speed in `local' and `online' learning and processing can be observed when P and D are integrated in CS TPNs and when P, D, and U are integrated in CS TPNs, compared to BDSP alone \textbf{(c)} BDSP XOR output reconstruction in 250 epochs \textbf{(d)} BDSP + P + D XOR output  in 250 epochs \textbf{(e)} BDSP + P + D + U XOR output reconstruction in 250 epochs. Note that cooperative CS TPNs + BDSP cross the target threshold faster with the same number of neurons.}
\vspace{-1.2em}
\end{figure*}

\begin{table*}[!htb]
	\caption{Training: Summary of the Main Equations. In the first column, $r_0$ is the receptive input and $y_l$ is the output of the TPNs which is modulated with context $c_l$ through the non-linear function. The BDSP algorithm adopts the bursting rate depending on the difference between the current bursting probability $P_l(t)$ and the previous probability $P_l(t-1)$. Although the learning equations remain the same as standard BDSP, the bursting probability and bursting rates use cooperative output hence the changes are sensitive to the contextual modulation. Note that the output neurons are not context-sensitive but only the hidden-layer neurons. Moreover, $Y_{b_l}$ is the feedback weight matrix.}
	\begin{center}
		\resizebox{\textwidth}{!}{
			\begin{tabular}{cccccc}
				\hline
                \textbf{CS-TPNs + BDSP}  &   TPNs + BDSP \cite{belghazi2018mutual}  & PNs + BP \cite{lecun2015deep}  \\ 
                \hline
                $r_0 = x$ & $e_0 = x$ & $a_0 = x$ \\
                ${y_l} = {r_l} + {c_l} {\odot} {(0.1 + |r_l|)} + {c_l} {I_u} {\odot} {(2 + |r_l|)}$ & $e_l = f_l(W_l e_{l-1})$ & $a_l = f_l(W_l a_{l-1})$  \\
                where $c_l$ is integrated context and $I_u$ is the synergistic component & & \\
                ${y_L} = f_L(W_{L}y_{(L-1)})$ & $e_L = f_l(W_L e_{L-1})$ & $a_L = f_L(W_L a_{L-1})$  \\        \hline
                $\overline{p}_L = p^{(0)}_L (1, 1, \ldots, 1)^T$ & $\overline{p}_L = p^{(0)}_L (1, 1, \ldots, 1)^T$ & $g_L = f'_l(V_L) \odot \nabla_{a_L} \mathcal{L} $ \\
                $p_L = \zeta(p_L - {h(y_L)}) \odot \nabla_{{y_L}} \mathcal{L}$ & $p_L = \zeta(p_L - h(e_L)) \odot \nabla_{e_L} \mathcal{L}$ & $g_l = f'_l(V_l) \odot (W_{l+1}^T g_{l+1})$ \\
                $\overline{u}_l = {h(y_l)} \odot (Y_{l}b_{l+1}) $ & $\overline{u}_l = h(e_l) \odot (Y_{l}b_{l+1}) $ &  \\
                $\overline{p}_l = \sigma(\beta \overline{u}_l + \alpha) $ & $\overline{p}_l = \sigma(\beta \overline{u}_l + \alpha) $ & \\
                $\overline{b}_l = \overline{p}_l \odot {y_l}$ & $\overline{b}_l = \overline{p}_l \odot e_l$ &  \\
                $u_l = {h(y_l)} \odot (Y_{b_l+1})$ &  $u_l = h(e_l) \odot (Y_{b_l+1})$ &  \\
                $p_l = \sigma(\beta u_l + \alpha)$ & $p_l = \sigma(\beta u_l + \alpha)$ &  \\
                $b_l = p_l \odot {y_l}$ &  $b_l = p_l \odot e_l$ &  \\   \hline
                $\Delta W_l = \eta \delta b_l {y_{l-1}^T}$ & $\Delta W_l = \eta \delta b_l e_{l-1}^T$ & $\Delta W_l = -\eta g_l a_{l-1}^T$ \\ 
		\end{tabular}}
		\label{t.metrics}
	\end{center}
\end{table*}

In the standard BDSP \cite{payeur2021burst}, the simple TPNs (Figs 1a and 1b) integrate the FF information at the soma to create an event (spike or burst) disregarding the contextual input. Once the event is generated, the feedback information (context) at the apical dendrites decides whether this already generated event should be a singlet (incoherent information) or a burst (coherent information). Although BDSP uses the apical zone to solve the `online' credit assignment problem, the processing is still driven by PNs. In short, the learning is inspired by TPNs, but the processing is not \cite{adeel2022context, adeel2022unlocking, adeel2024cellular}. \\
In contrast, our approach shows that complex CS-TPNs (Figs 1c and 1d) at the soma flexibly integrate moment-by-moment rich contextual current (including FB, proximal, distal, and thalamus (universal)) with the FF somatic current such that the somatic current is amplified when both FF and context are coherent; otherwise, it is attenuated. This generates the event only when the somatic and apical currents are coherent, which is then translated into a singlet or a burst based on the FB information. Simulation results show that this flexible integration of somatic and contextual currents requires a reduced number of overall events (both singlets and bursts) in the system to generate external stimuli (Fig 1d). Raster plots shown in the later sections show that CS-TPNs burst far more (transmitting coherent information) than singlets (transmitting incoherent information), hence they learn faster.\\
The local weights are updated based on local burst firing information. The bursting (y) is controlled by a novel asynchronous modulatory function (MOD) \cite{adeel2022context, adeel2022unlocking} that ensures neurons burst only when R is coherent with C, leading to the suppression of conflicting information (singlets), and simultaneous faster processing and learning \cite{adeel2022context, adeel2022unlocking}. The updated BDSP learning rule is detailed in Table 1 and compared with standard BDSP and backpropagation (BP). Adapted from \cite{payeur2021burst}, somatic membrane potential dynamics and the apical dendrite dynamics of CS-TPNs are represented by the following simplified differential equations (see Eqs: \ref{Vsoma} - \ref{Mod2}):


\begin{equation}
    \dot{V_s} = \frac{1}{\tau_{s}}(V_s - E_L) + \frac{1}{C_s}Mod(I_{s},I_{c}) - \frac{1}{C_s}w_{s}
    \label{Vsoma}
\end{equation}
\begin{equation}
    \dot{w_s} = - \frac{1}{\tau_{ws}}(w_s) + bS(t)
    \label{wsoma}
\end{equation}


The MOD function is capturing the complex interactions between somatic and dendritic currents moment-by-moment at the soma, enabling intrinsic adaptation mechanism, allowing nuanced control over the neuron’s firing rate and pattern. \\
$I_C$, including $I_P$, $I_D$, and $I_U$, is solely integrated at the apical site using the TPN dynamics defined in \cite{payeur2021burst}. The network evolves on the same time scale, as coincidental detection is key in two-point neuron operations \cite{larkum1999new, aru2020apical}. Since $I_U$ is the most reliable contextual field, constituting the coherent/synergistic signals across the MS network and acquired by first conditionally segregating the coherent and incoherent MS information streams and then recombining only the coherent multistream, it has been awarded the maximum weight in the integration function. To make it equivalent to G (coupling gate between apical and basal sites) (Aru et al., TICS, 2020), its influence is embedded within the MOD function at the soma, such that terms with Iu dominate the overall coupling between $I_S$ and $I_S$. The third term tends to zero in the absence of $I_U$. The offset value of 2 is empirically calculated. Nevertheless, there could certainly be a better way to model the whole dynamics. The MOD functions are explicitly defined in the later sections.

Where $V_s$ represents the membrane potential of the somatic compartment (soma), $\tau_s=16ms$ is the membrane time constant, $C_s$=370pF is the membrane capacitance, $E_L$=-70mV is leak reversal potential, and $w_s$ is an adaptation variable. The total current applied to the soma is represented by $Mod(I_s, I_{c}, I_{u})$ that is the basal current $I_s$ modulated with the contextual current $I_{c}$ and the synergistic components $I_{u}$. The basal current here is the integrated synaptic inputs (i.e., receptive excitatory and inhibitory inputs) along with basal noise. The contextual current is the accumulated effect of proximal signal, distal signal, and feedback error with dendritic noise. Whereas, the synergistic components $I_{u}$ is the extracted synergistic FF components. 
The adaptation variable $w_s$ is defined by the \eqref{wsoma} where $S(t)$ is the spike train of the neurons, and $b=200$ is the strength of spike-triggered adaptation. A spike occurs, every time $V_s$ crosses a dynamic threshold (i.e., $-50mV$). Subsequently, the threshold increases by $2mV$ immediately following a spike and returns to $-50mV$ with a time constant of $27ms$. After a spike occurs, the $V_s$ is reset to a resting voltage $V_r=-70mV$. These values of the parameters are adopted from the \cite{payeur2021burst}.

The CS-TPNs model receives the contextual signals (i.e., FB (E), P, D, and U) at the apical dendritic site and integrates them individually. Their respective membrane potentials, termed $V_e$, $V_p$, $V_d$, and $V_u$ are defined by the following simplified differential equations adopted from \cite{payeur2021burst}:

\begin{equation}
    E = \dot{V_e} = \frac{1}{\tau_e}(V_e - E_e) + \frac{1}{C_e}I_e - \frac{1}{C_e}w_e
    \label{Vdend}
\end{equation}
\begin{equation}
    \dot{w_e} = - \frac{1}{\tau_{w_e}}w_e + \frac{1}{\tau_{w_e}}a_w(V_e - E_L)
    \label{Wdend}
\end{equation}

\begin{equation}
    P = \dot{V_p} = \frac{1}{\tau_p}(V_p - E_p) + \frac{1}{C_p}I_p - \frac{1}{C_p}w_p
    \label{Vgossip}
\end{equation}
\begin{equation}
    \dot{w_p} = - \frac{1}{\tau_{w_p}}w_p + \frac{1}{\tau_{w_p}}a_w(V_p - E_L)
    \label{Wg}
\end{equation}

\begin{equation}
    D = \dot{V_d} = \frac{1}{\tau_d}(V_d - E_d) + \frac{1}{C_d}I_d - \frac{1}{C_d}w_d
    \label{Vcross}
\end{equation}
\begin{equation}
    \dot{w_d} = - \frac{1}{\tau_{w_d}}w_d + \frac{1}{\tau_{w_d}}a_w(V_d - E_L)
    \label{Wcross}
\end{equation}

\begin{equation}
    I_c = \frac{g}{C}(f(V_p) + f(V_e) - f(V_d))
    \label{Icontext}
\end{equation}

\begin{equation}
    U = \dot{V_u} = \frac{1}{\tau_u}(V_u - E_u) + \frac{1}{C_u}I - \frac{1}{C_u}w_u
    \label{Vmemory}
\end{equation}
\begin{equation}
    \dot{w_u} = - \frac{1}{\tau_{w_u}}w_u + \frac{1}{\tau_{w_u}}a_w(V_u - E_L)
    \label{Wu}
\end{equation}

\begin{equation}
    I_u = \frac{g}{C}f(V_u)
    \label{Ithalamus}
\end{equation}

In the aforementioned equations, $\tau_d$, $\tau_g$, $\tau_c$ are membrane time constants ($7ms$ each). $C_d$, $C_g$, and $C_c$ are membrane capacitance ($170pF$ each). Similarly, the $\tau_{w_{d}}$, $\tau_{w_{g}}$, and $\tau_{w_{c}}$ are adaptation time constants ($30ms$ each). $\alpha_w = 13nS$ is sub-threshold adaptation. 
$g=1,300 pA$ in the eq\eqref{Icontext} is the dendrosomatic coupling. 
The simulation of the proposed model uses the same settings as \cite{payeur2021burst} including but not limited to dendrite-targeting inhibition, perisomatic inhibitions, noise, and synapses.
The top-down effect of the apical compartment is $I_{c}$, represented by eq\eqref{Icontext} and $I_{u}$ in eq\eqref{Ithalamus} is the gating current referring to synergistic components.


Contrary to the multiplexing approach \cite{naud2018sparse} used in BDSP \cite{payeur2021burst}, the contextual modulatory function $Mod(I_s, I_{c})$ detailed in \cite{adeel2020conscious, adeel2023unlocking} is scaled as follows to suit the spiking neural net (SNN) in the CS-TPNs circuit.

\begin{equation}
    Mod(I_s, I_{c}) = I_s + I_{c} (0.1 + |I_s|)
    \label{Mod}
\end{equation}

The scaling is necessary to align with the working range values and the scaling is calculated empirically. This however, keeps the essence of the contextual field (C) overruling the typical dominance of the receptive field (R), and therefore discourages and encourages amplification of neural activity when C is weak and strong, respectively \cite{adeel2020conscious, adeel2023unlocking}.
As discussed earlier, to incorporate the synergistic components U as the Gate, the modulatory function is further modified (see Eq \ref{Mod2}).

\begin{equation}
    Mod(I_s, I_{c}, I_{u}) = I_s + I_{c} (0.1 + |I_s|) + I_{c} I_{u}(2+ |I_s|)
    \label{Mod2}
\end{equation}

In the Eq \ref{Mod2}, the gating current is $I_{u}$. In this modified modulatory function, the third term incorporates both `U' and the integrated context `C' with higher weight than the second term. In the presence of U, the third term dominates the overall results, whereas the third term tends to zero in the absence of U. The offset value of $2$ is empirically calculated.

\section{Results}
\label{s.results}

\subsection{Shallow spiking XOR}

\begin{figure*} 
	\centering
	\includegraphics[trim=0cm 0cm 0cm 0cm, clip=true, width=1\textwidth]{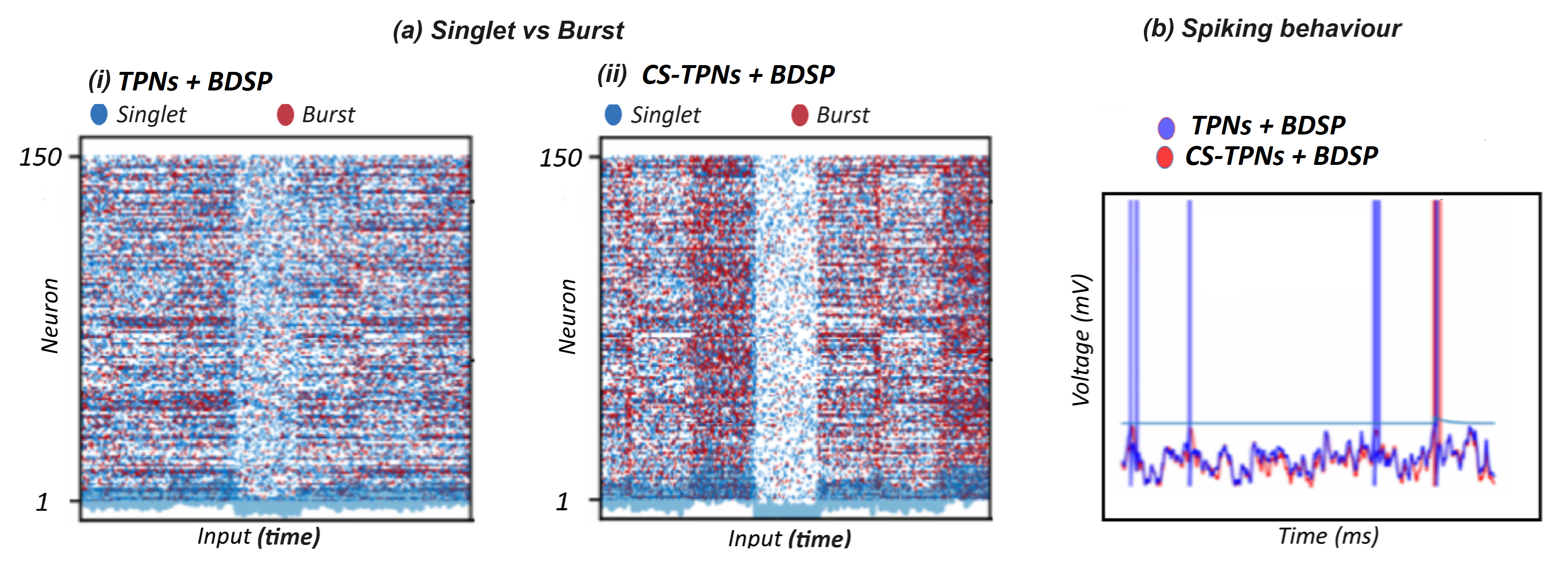}
	\caption{(a) Raster plots with 150 neurons: TPNs + BDSP (i) cooperative CS-TPNs + BDSP (ii).  CS-TPNs tend to remain largely silent when information is less relevant but become active (bursting) when information is relevant, compared to BDSP alone. Note the clear distinction between coherent (burst) and conflicting signals (singlet). It implies that CS-TPNs are bursting far more (transmitting coherent information) than singlets (transmitting incoherent information), hence learn faster. The bottom blue line reflect event rates. (b) it is notable that TPNs fire more frequently than CS-TPNs. (c) A 50-layer deep CNN composed of CS-TPNs requires significantly fewer neurons at any time during training with better generalization capability (Table II) compared to a deep CNN composed of PNs. This reveals the universal applicability of CS-TPN-driven efficient information processing regardless of the learning mechanism type. See: http://cmilab.org/research/}
\vspace{-1.2em}
\end{figure*}

To simulate the spiking CS-TPNs trained using BDSP, we adopt a 3-layer network structure presented in \cite{payeur2021burst} with similar settings for exclusive or (XOR) tasks (Fig. 2a). The proposed network consists of two neural pools (termed ensemble) at the input, two ensembles in the hidden layer and one ensemble at the output. Each ensemble has a fixed population of interconnected CS-TPNs. To perform an XOR function, the network must respond with a high output if only one input is active, and a low output if neither or both input pools are active. The XOR is a classical non-linear example to demonstrate the capability of a network. The network is initialized such that the output ensemble treats any input combination as roughly equivalent. 


For analysis, the network performance and neural activity of CS-TPNs is compared to a similar network but with context-insensitive TPNs. The cost function for the network is binary cross entropy. The population size in each ensemble for BDSP only and for CS-TPNs+BDSP network is 175 neurons. The total population of these two networks is equal (i.e., 875 neurons). The CS-TPNs+BDSP network with U (CSM-TPNs+BDSP) has an additional ensemble representing U with a population of 50 PNs. To retain the network size, the input, hidden and output ensemble population size is reduced to 150 neurons each. Thus the total population size of the CSM-TPNs+BDSP network is 800. 
Fig. 2b shows the convergence of the cost function over 250 epochs for each network. The cost functions are compared to demonstrate the learning speed and convergence of BDSP alone (blue), CS-TPNs+BDSP (orange), and CSM-TPNs+BDSP (green). It can be seen that BDSP with CS TPNs learns faster than the BDSP, whereas, the BDSP with CSM-TPNs learns even quicker. The network output is shown in Fig. 2c,d,e. See the region where the mean of the function at (0, 1) and (1,1) inputs is distant from the threshold in CS-TPNs+BDSP and CSM-TPNs+BDSP (Fig. 2d, e) compared to BDSP alone (Fig. 2c). Fig. 3a shows the raster plots of standard BDSP and CS-TPNs+BDSP. The blue dots represent the event rate and the red dots represent the bursts. Higher bursts are associated with the high output whereas, the lesser events and bursts represent low output. It can be seen that the network with CS-TPNs distinguishes between the high and the low outputs more clearly than BDSP alone CS-TPNs. Specifically, it is to be observed that CS-TPNs tend to remain largely silent when information is less relevant (close to zero) but become active (bursting) when information is relevant (close to one). It also implies that CS-TPNs burst more when apical and basal, both the inputs are stronger, thus amplifying and suppressing the transmission of coherent and conflicting information, respectively. This is because the neighbouring proximal TPNs and distal TPNs in the same and other ensembles influence the perception of each TPN in each ensemble. They are sharing their perceptions with each other with the goal to minimise the conflicting information and maximise the coherent information, achieving harmony across the network. Additionally, it is notable that context-insensitive TPNs in BDSP burst more frequently than CS-TPNs, implying that neurons are not sharing information and burst even when the received information is not very relevant to the task at hand. Fig. 3b shows the development of the membrane potential of a single randomly selected neuron. During a given time frame, the context-insensitive TPNs in BDSP fire three times more than CS-TPNs. The CS-TPNs fire only when the proximal TPNs and the distal TPNs emphasize the need to fire, thus quickly distinguishing between the irrelevant and relevant events for the task at hand. It can be convincingly concluded that CS-TPNs are more robust and efficient, capable of learning with a smaller population of neurons compared to the context-insensitive TPNs in BDSP.\\
The Fig 1a shows the results reproduced from the BDSP which demonstrates that the BDSP with short-term plasticity supports multiplexing of FF and FB signals. The graphs also reflect the signal-processing behaviour of the neuronal population. Fig. 1a shows a conventional TPNs circuit and Fig. 1b is a circuit consisting of CS-TPNs. The circuit architecture remains similar in both cases, having two populations of TPNs (Pop 1 and Pop 2) and two populations of PNs (disk and square). The PNs population provides inhibition. The external input to pop 1 is somatic current ($I_s$) and that to pop 2 is dendritic current ($I_d$). The output of Pop 1 is somatic input for Pop 2 whereas, the Pop 2 output is fed to the apical at Pop 1 as feedback. For simplicity, in the CS-TPNs network (Fig. 1b), the D and U are set to zero and only P is connected. The top two graphs in Fig. 1a, b reflect the activity in Pop 2, i.e., the burst probability of the Pop 2 neurons against the applied current $I_d$ and event rate against the FF input from Pop 1, respectively. Similarly, the two bottom graphs in Figs 1a and 1b represent the activity at Pop 1. These graphs show the burst probability at pop 1 due to the feedback from pop 2 and the event rate against the applied current $I_s$. 
The modulatory function of CS-TPNs results in dynamic and variable neural activity with an increase in burst probability and event rate compared to conventional TPNs. This suggests a higher responsive system influenced by contextual inputs while retaining the multiplexing behaviour of BDSP.

\begin{table*}[!htb]
	\caption{Neural activity, PESQ, and STOI Results: The Ideal column represents the reference value that does not apply to neural activity.}
        \label{tab:tfaccuracy}
	\begin{center}
		\resizebox{\textwidth}{!}{
			\begin{tabular}{c|ccc|ccc|ccc|}
 			\toprule
 			\multicolumn{1}{c}{SNR} &
    		\multicolumn{3}{c|}{Neural Activity} &
 			\multicolumn{3}{c|}{PESQ} &
 			\multicolumn{3}{c}{STOI} \\  
 			\midrule
 			&
 			\multicolumn{1}{c}{Ideal} &
 			\multicolumn{1}{c}{Baseline} &
 			\multicolumn{1}{c|}{\textbf{Proposed}} &
 			\multicolumn{1}{c}{Ideal} &
 			\multicolumn{1}{c}{Baseline} &
 			\multicolumn{1}{c|}{\textbf{Proposed}} &
   			\multicolumn{1}{c}{Ideal} &
 			\multicolumn{1}{c}{Baseline} &
 			\multicolumn{1}{c}{\textbf{Proposed}} \\
 			\midrule
 			-6dB &  & 0.34316 & 0.00102 & 1.92 &	1.23 & 1.33 & 0.78 &	0.56 &	0.63 \\  
 			0dB &  & 0.31665 & 0.00102 & 2.09 &	1.40 & 1.58 & 0.83 & 0.63 & 0.71 \\
 			6dB & &	0.33646	& 0.00102 & 2.28 & 1.52 & 1.77	& 0.87 & 0.71 & 0.78 \\		
 			\bottomrule
		\end{tabular}}
		\label{t.metrics_1}
	\end{center}
\end{table*}

\begin{table*}[!htb]
	\caption{Sparsity Ratio (SR) and MACs comparison: PNs vs. TPNs.  Given a non-zero propagation only, the neural activity can be interpreted in terms of the reduction of multiplications and accumulations (MACs) which results in floating point operations (FLOPs). The MACs are calculated based on the number of convolutions used and the modulation function adopted in context-sensitive networks. The sparsity ratio is chosen from 0dB SNR. The after-training MACs are the inference MACs. The context-sensitive network has $24\%$ more parameters due to context-extracting layers and modulation. This is visible with more MACs (i.e., 1.394 times more MACs for context-sensitive networks). However, the overall reduction in MACs after training is 980 times for context-sensitive networks compared to 3.16 for the standard network   }
	\begin{center}
		\resizebox{\textwidth}{!}{
			\begin{tabular}{cccccc}
				\toprule
				 Model & Sparsity Ratio (SR) @ 0dB & MACs & After training MACs & MACs Reduction (times) \\
				\midrule
				{Baseline-(PNs)} & 0.31665 & 36827136 & 1161313 & 3.16 \\ 
				\midrule
				{\textbf{CS-TPNs}} & 0.00102 & 51342336 & 52369 & 980.39 \\ 
				\bottomrule
		\end{tabular}}
		\label{t.metrics_2}
	\end{center}
\end{table*}

\begin{table*}[!htb]
	\caption{FLOPs comparison: PNs vs. CS-TPNs}
	\begin{center}
		\resizebox{\textwidth}{!}{
			\begin{tabular}{cccccc}
				\toprule
				 Model &  @ 0dB & FLOPs & After training FLOPs & FLOPs Reduction (times) \\
				\midrule
				{Baseline-(PNs)} & 0.31665 & 73654272 & 23322625 & 3.16 \\ 
				\midrule			{\textbf{CS-TPNs}} & 0.00102 & 102684672 & 104738 & 980.39 \\ 
				\bottomrule
		\end{tabular}}
		\label{t.metrics_3}
	\end{center}
\end{table*}

\subsection{Deep AV speech processing with 50-layered CNN composed of CS-TPNs}
\subsubsection{AV Dataset for Speech Enhancement \cite{adeel2023unlocking}}
The AV ChiME3 dataset was created by blending the clean Grid videos \cite{cooke2006audio} with ChiME3 background noises \cite{barker2017third} (such as cafe, street junction, public transport (BUS), and pedestrian areas) across SNR levels from -6 to 6dB, using a 3dB increment. The preprocessing steps include adding prior visual frames to the data. To capture temporal dynamics, the system utilizes six preceding frames of both audio and visual data, enhancing the correlation between visual and auditory features. The Grid corpus includes recordings from 34 speakers, with each speaker delivering 1000 sentences. From this group, a subset of 4 speakers was chosen (comprising two white females and two white males), each contributing 900 command sentences. This selection aims to maintain speaker diversity. Additional information is detailed in \cite{adeel2022context, adeel2021contextual}. For training and evaluation, the data is divided into a 75\%-25\% split, with one sample reserved from both the training and testing portions as a representative proxy.

\begin{figure*} 
	\centering
	\includegraphics[trim=0cm 0cm 0cm 0cm, clip=true, width=0.65\textwidth]{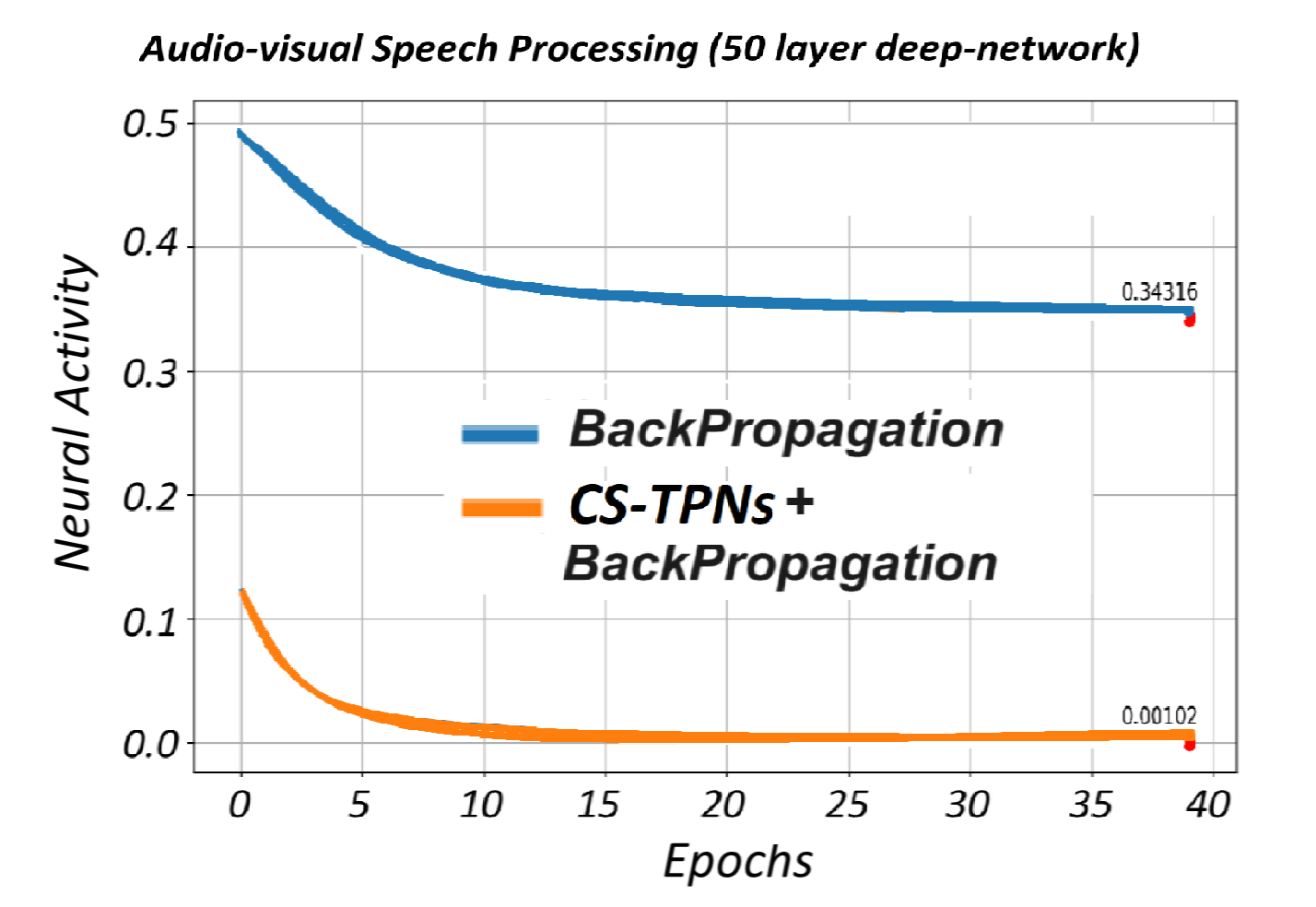}
	\caption{A 50-layer deep CNN composed of CS-TPNs requires significantly fewer neurons at any time during training with better generalization capability (Table II) compared to a deep CNN composed of PNs. This reveals the universal applicability of CS-TPN-driven efficient information processing regardless of the learning mechanism type.}
\vspace{-1.2em}
\end{figure*}

\subsubsection{Network Architecture:} The two-layered CNN composed of CS-TPNs \cite{adeel2022context} is scaled to a 50-layered CNN. Each TPN in each convolutional layer used the novel MOD block. The Mod block utilizes a structure and a modulatory transfer function as described in \cite{adeel2022context, adeel2023unlocking}, adjusting its scale based on the input values range. The multi-modal AV network architecture integrates two parallel streams for audio and visual modalities. The audio modality processes the magnitude of the noisy speech's Short-Time Fourier Transform (STFT), while the visual modality processes a series of images through 2D convolution layers to align with the audio stream's dimension. These deep models are tasked with reconstructing a clean STFT of the audio signal from both noisy audio and visual inputs as mentioned in \cite{gogate2020cochleanet}. In each stream, the data passes through convolution units that serve as feature extraction layers, each followed by a ReLU unit, a contextual modulation, constituting a network block.  Each modality employs 12 such blocks, with each containing two convolution units and following the aforementioned sequence including context modulation. The context is derived exclusively from P and D, where the audio stream, D is represented by visuals and P by the audio itself, and vice versa for the visual stream. Throughout these blocks, both audio and visual signals undergo frequency domain reduction while maintaining the time domain integrity. The outputs from the final blocks of both streams are then averaged, pooled, concatenated, and fed into a dense layer followed by a sigmoid activation function. The network employs convolution layers with 16 filters and a kernel size of 3, and it is trained end-to-end using back-propagation. The specific loss function employed for this model is detailed in \cite{adeel2022context}
The MOD function is expressed as:
\begin{equation}
    Mod(R, C) = \zeta[R*C*(R^2 + 2*R + C*(1+|R|))]
    \label{Moddeep}
\end{equation}
Where $\zeta$ is the activation function, ReLU6 in this case.

The Loss function is expressed as:

\begin{equation}
    Loss = \beta E [SE(Z, \hat{Z})] + \gamma E[\epsilon]
    \label{Mod_eq}
\end{equation}
The $SE$ is the squared error between the clean speech $Z$ and the predicted speech $\hat{Z}$. However, the second term with $\epsilon$ is a differentiable approximation for the number of non-zero activations'. The coefficients of the loss function are adjusted to make the secondary objective significantly less important than the main goal; in particular, $\gamma$ is set to a really small value in all experiments. The model estimates the clean spectrogram when the noisy spectrogram and visual images are fed as input.

Table 1 compares the standard PNs-inspired CNN (baseline) with the proposed CS-TPNs inspired CNN in terms of neural activity, perceptual evaluation of speech quality (PESQ), and short-time objective intelligibility (STOI). It can be seen (see Fig. 4) that the CS-TPNs inspired CNN generalises better than baseline with up to 330x fewer neurons for all SNRs. Same proportion could be observed in required number of MACs (Multiply-Accumulate Operations) and FLOPs (Floating Point Operations).

\section{Conclusion}
\label{s.discussions}
The reliance on PNs and BP continues. Despite their remarkable performance improvements in a range of real-world applications, including large language models (LLMs), current AI technology based on them is rapidly becoming economically, technically, and environmentally unsustainable. The limitations of current AI systems therefore persist: issues with scalability, slow and unstable training, processing of irrelevant information, and catastrophic forgetting—all requiring a massive number of electronic components, which leads to trade-offs between cost, speed, power, size, accuracy, generalization, and inter-chip congestion. CMI-inspired CS-TPNs driven deep networks proposed here have shown the capability of processing large amounts of heterogeneous real-world AV data using a significantly smaller number of neurons compared to standard PNs-inspired deep nets, with better generalization in some cases. The demonstration of biologically plausible CS-TPNs simulation with BDSP shows that this efficient information processing approach is universal regardless of the learning algorithm. Future work includes further scaling up the proposed artificial and spiking TPNs inspired neural nets for a range of different real-world problems.





\section{Acknowledgments}
\label{s.acknowledgment}

This research was supported by the UK Engineering and Physical Sciences Research Council (EPSRC) Grant Ref. EP/T021063/1. We would like to acknowledge Professor Bill Phillips and Professor Leslie Smith from the University of Stirling, Professor Newton Howard from Oxford Computational Neuroscience,  Professor Heiko Neumann from the Institute of Neural Information Processing, Ulm University, Professor Panayiota Poirazi from IMBB-FORTH and Dr. Michalis Pagkalos from the University of Crete, Foundation for Research and Technology Hellas for their help and support in several different ways, including reviewing this work, appreciation, and encouragement. \\ \\
\textbf{Code availability:} The data and code supporting the findings of this study are available upon request.



\bibliographystyle{IEEEtran}
\bibliography{references}

\section{supplementary material}
\label{s.appendix}


\begin{figure*} 
	\centering
	\includegraphics[trim=0cm 0cm 0cm 0cm, clip=true, width=1\textwidth]{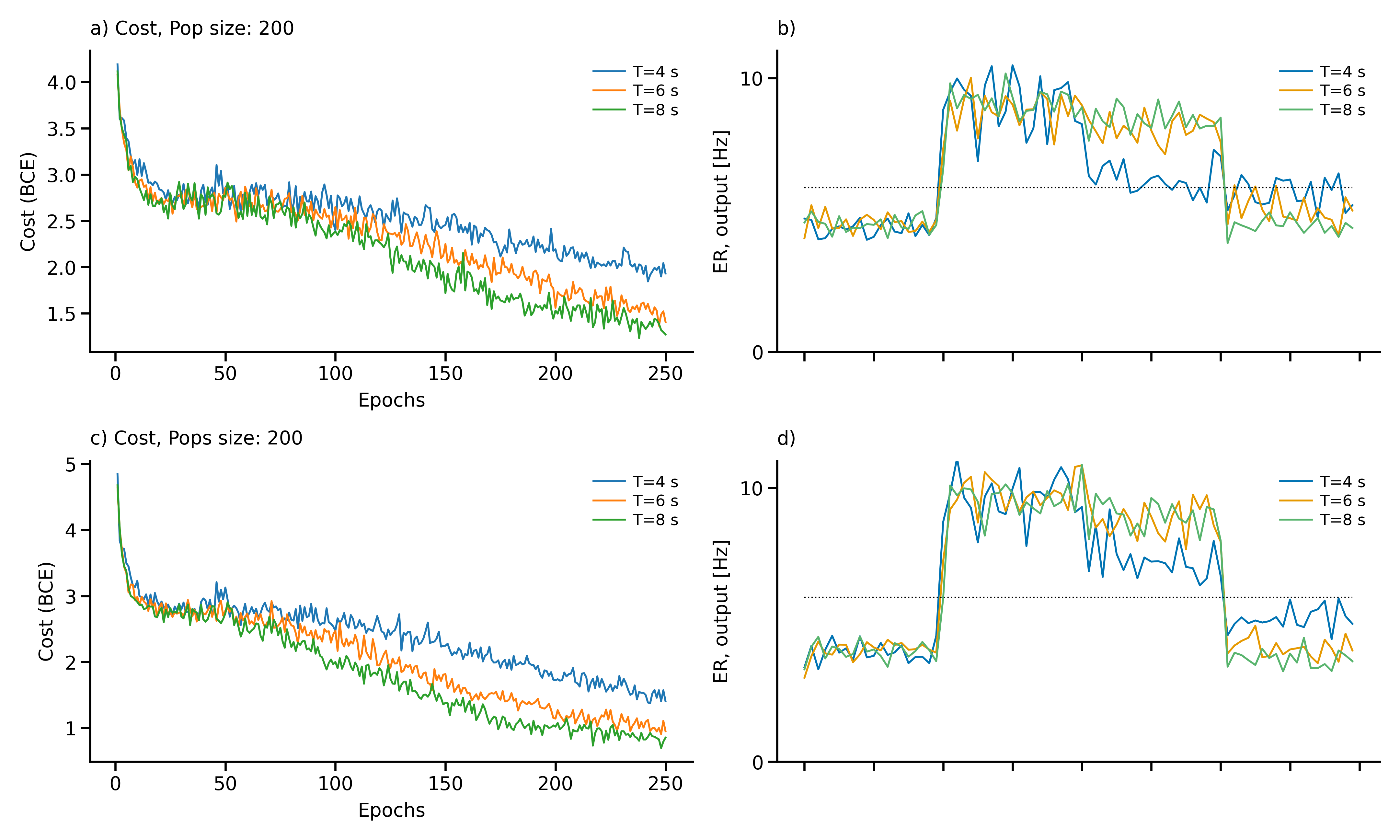}
	\caption{\textbf{Effects of different time scales on the XOR task for the population size of 200 neurons on Standard BDSP and CS-TPNs+BDSP.} \textbf{a} and \textbf{c}, Comparison of costs for different duration of examples T (in s). T = 8s (green line) is the duration used in Fig. 2. \textbf{b} and \textbf{d}, Output event rate (ER) after learning for the cases in \textbf{a},\textbf{c} classifying between 'true(1)' and 'false(o)' for the XOR. The costs for the standard network (\textbf{a}) reach a minimum of $1.5$ starting at epoch 250. The context-sensitive network attains the cost of $1.0$ starting at epoch $120$. It is evident from the graphs that the FF context-sensitive network learns faster and better.
 }
\vspace{-1.2em}
\end{figure*}

\end{document}